# Magnetism of σ-phase Fe-Mo alloys: its revealing and characterization

J. Cieślak[1*], S. M. Dubiel[1] and M. Reissner[2]

[1]AGH University of Science and Technology, Faculty of Physics and Applied Computer Science, PL-30-059 Krakow, Poland

[2]Institute of Solid State Physics, Vienna University of Technology, A-1040 Wien, Austria

**Abstract**

A low-temperature magnetism was revealed in a series of σ-$Fe_{100-x}Mo_x$ alloys ($x$=45-53). Its characterization has been done using vibrating sample magnetometry, Mössbauer spectroscopy, and ac magnetic susceptibility. The magnetic ordering temperature was determined to lie in the range of ~46 K for $x$=45 and ~22K for $x$=53, and the ground magnetic state was found to be typical of a spin-glass.

PACS: 75.30.Cr, 75.30.Kz, 75.50.Ba, 75.50.Lk

The sigma (σ) phase (space group $D_{4h}^{14}$ - $P4_2/mnm$) is one of many members of the Frank-Kasper (FK) phases [1]. It can be formed in alloys in which at least one constituting element is a transition metal. The tetragonal unit cell of σ-phase hosts 30 atoms distributed over 5 different lattice sites, usually termed as A, B, C, D and E. The sites have high coordination numbers (12-16), and the distribution of atoms is not stoichiometric. These features, in combination with the fact that σ-phase can be formed in a certain range of composition, are the reason that σ-phase alloys show a diversity of physical properties that can be tailored by changing constituting elements and/or their relative concentration. Structural complexity and chemical disorder make them also an attractive yet challenging subject for investigation. The interest in σ-phase is further justified by a deteriorating effect on useful properties of technologically important materials in which it precipitates [2, 3]. It should be, however, added that attempts have been undertaken to profit from its high hardness and to use profit from this feature by increase of materials strength e. g. [4, 5].

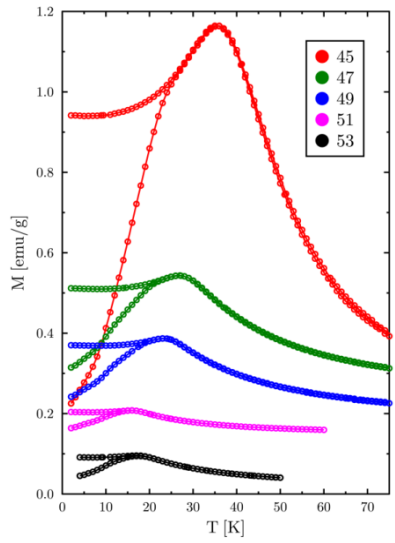

Fig. 1 (Color online) Field cooled and zero-field cooled magnetization curves of the studied samples recorded in external magnetic field of 100 Oe vs. temperature, *T*. The legend shows the content of Mo in at%.

Concerning magnetic properties of σ-phase in binary alloys, so far only σ-phase in Fe-Cr, Fe-V and Fe-Re systems was definitely evidenced to be magnetic [6-9]. Its magnetism has been recently shown to be more complex than initially anticipated viz. it has a spin glass structure and at least in former two cases a re-entrant character [8, 9].

Based on a systematic experimental (vibrating sample magnetometry, VSM, $^{57}$Fe Mössbauer spectroscopy, MS, and ac susceptibility) measurements a clear evidence was found that σ-FeMo alloys are also magnetic, and their magnetism is of a spin-glass (SG) type, too.

A series of five σ-$Fe_{100-x}Mo_x$ (*x*=45-53) sample was prepared in the following way: powders of elemental iron (3N+ purity) and molybdenum (4N purity) were mixed together in appropriate proportions and masses (2g), and next pressed to pallets. The pellets were subsequently isothermally annealed at 1703 K during 6 h, and afterward quenched into liquid nitrogen. The mass loses of the fabricated samples were less than 0.1% of their initial values, so it is reasonable to take their nominal compositions as real ones. X-ray diffraction patterns recorded on the powdered samples gave evidence that their crystallographic structure was that of σ. More details regarding sample preparation and verification can be found elsewhere [11]. Measurements of magnetization, *M*, were performed using a vibrating sample magnetometer versus temperature (in a constant magnetic field), and versus external magnetic field at different temperatures. *M(T)*-curves were recorded in an external magnetic field $\mu_o H$=10 mT in field cooled (FC) and zero-field cooled (ZFC) regimes. The measured *M(T)* magnetization curves are displayed in Fig. 1, while the *M(H)*-curves are displayed in Fig. 2.

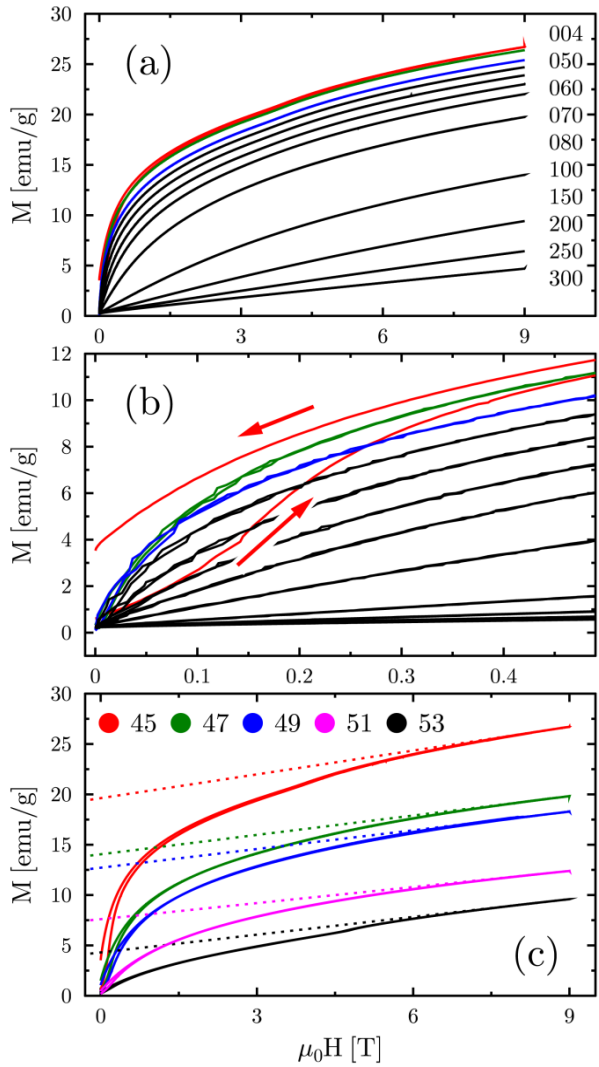

Fig. 2 (Color online) Magnetization curves measured at 4K as a function of an external magnetic field, $\mu_oH$. (a) Curves recorded on the σ-$Fe_{53}Mo_{47}$ sample measured at various temperatures (indicated on right) (b) the same as (a) but in the narrower range of magnetic field. (c) Curves for all investigated compounds measured at 4K. Dotted lines represent extrapolation of high-field linear parts of the *M*-curves. The legend shows the content of Mo in at%.

Preliminary measurements of the ac magnetic susceptibility were performed on the σ-$Fe_{49}Mo_{51}$ sample versus frequency ranging between 10Hz and 10k Hz. As one can clearly see in Fig. 3, χ'(T) curve has a cusp that shifts with frequency to higher values of temperature, a feature characteristic of spin glasses.

Fig. 3 (color online) Real part of the ac magnetic susceptibility vs. temperature and frequency.

The $^{57}$Fe Mössbauer spectra were recorded on powdered samples with a density of ~10 mg Fe/cm$^2$ in the temperature range of 6-80 K using a standard spectrometer with a sinusoidal drive and a $^{57}$Co/Rh source for the 14.4 keV gamma rays. Examples are shown in Fig. 4a.

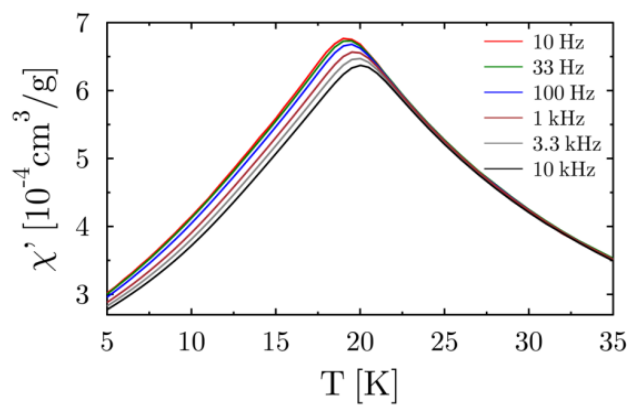

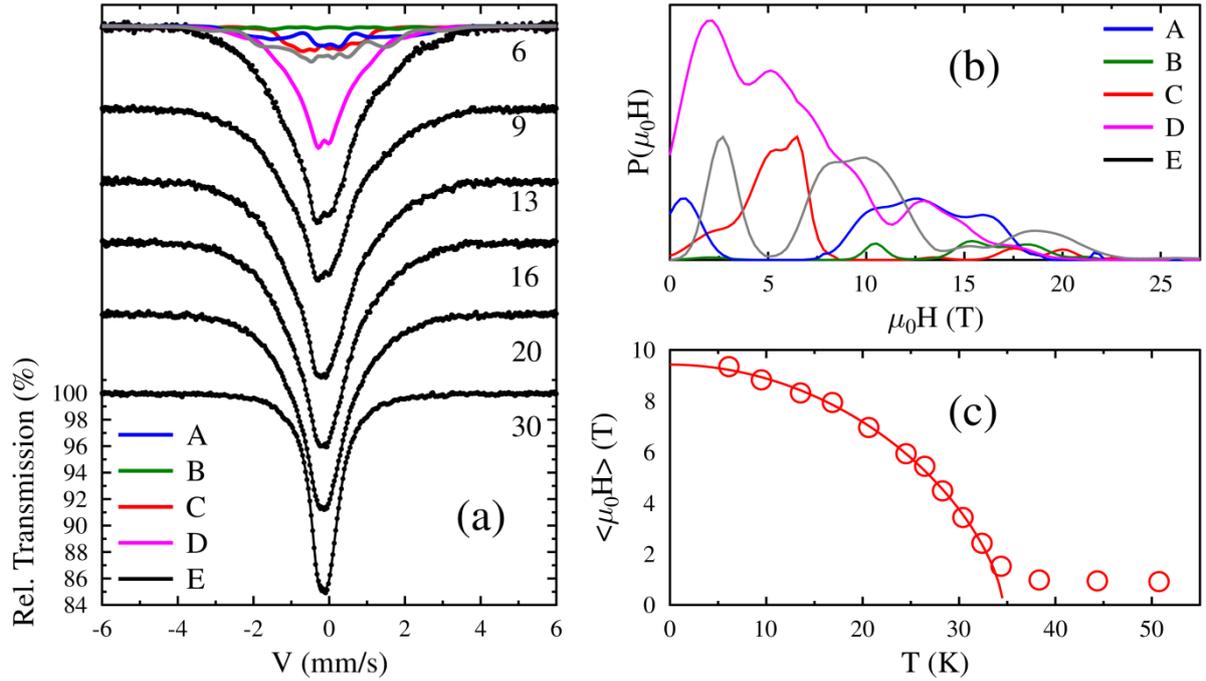

Fig. 4. (Color online) (a) $^{57}$Fe Mössbauer spectra recorded on σ-Fe$_{53}$Mo$_{47}$ sample at various temperatures, $T$, shown, (b) HFD-curves for the 6K spectrum, and (c) average hyperfine field as determined from the spectra and plotted vs $T$. Solid line is a guide to the eye.

A broadening of the spectra on lowering temperature can be clearly seen. In the light of the magnetization measurements it has evidently a magnetic origin. The hyperfine magnetic field, $H$, can be regarded as the quantity of merit, though its relationship to the magnetic moment, μ, is, in general, not linear [12]. In any case, the temperature dependence of the average hyperfine field, $<H>$, can yield a magnetic ordering temperature. The spectra, even at the lowest temperature, are not well-resolved what, on one hand, is due to the weak magnetism of the studied samples, and on the other hand to their complex crystallographic structure and chemical disorder. It is obvious that one cannot uniquely analyze the spectra in terms of five sub spectra corresponding to the five sub lattices present in the unit cell of σ. Instead, the spectra were analyzed in terms of electronic structure calculation-aided superposition of five hyperfine field distributions, HFD, taking also into account the quadrupole interactions. In particular, it was assumed that the charge-density and the electric field gradient obtained from the electronic structure calculations using charge and spin self-consistent Korringa–Kohn–Rostoker (KKR) Green's function method [13] are proportional to the isomer shift and the quadrupole splitting, respectively. Additionally relative amounts of individual sub spectra, $A_i$ were proportional to the iron occupancy of $i$-th sub lattice in the unit cell (known from XRD measurements). All mentioned parameters were treated as constant, and the shape of the individual distribution was constructed as a sum of 10 Gaussians. An example of the HFD-curves is illustrated in Fig. 4b. The $<H_i>$-values were obtained by integration the HFD-curves for the $i$-th sub lattice. The knowledge of the latter enabled the calculation of the total average value of the hyperfine field, $<H> = \sum <H_i> A_i$. An example of the $<H>=f(T)$ dependence is shown in Fig. 4c.

The data shown in Fig. 1 give clear evidence that all investigated samples are magnetic, and they show irreversibility. Such behavior is a characteristic feature of spin-glasses (SG), as recently revealed for the σ-phase in Fe-Cr, Fe-V and Fe-Re intermetallic compounds [6-9]. In the case of the σ-FeMo samples, two characteristic temperatures in the $M(T)$-curves can be defined viz. a temperature, at which the FC and ZFC curves go apart from each other known as the spin-freezing temperature, $T_{ir}$, and a second one, $T_m$, at which the ZFC-curve has its maximum. $T_{ir}$ is usually interpreted as the temperature of a transition into the SG state, whereas $T_m$ as the transition temperature into a strong irreversibility regime within SG. However, in contrast to the previously investigated σ-phase Fe-Cr, Fe-V and Fe-Re alloys, in the present case $T_{ir}$ is lower than $T_m$. Furthermore, both $M(T)$-curves are indistinguishable from each other for $T \geq T_m$ what might suggest that $T_m$ is the Néel temperature i.e. there is an antiferromagnetic (AFM) ordering for $T_{ir} \leq T \leq T_m$. An PM→AFM →SG transition would not be exotic as such transitions were found previously e. g. [11-13]. However, we have evidence – see Fig. 5 - that a magnetic ordering occurs already at $T_C > T_m$ signaling thereby the possibility of two different ordered magnetic states prior to the ground SG state. This puzzling issue cannot be uniquely solved based on the presently performed measurements. Hopefully, neutron diffraction measurements we are planning to carry out will help to clear the type of ordering that evidently occurs between the PM and SG states. The SG-character of the magnetism in the σ-FeMo alloys and the deteriorating effect of Mo can be also inferred from the measurements presented in Fig. 2, where two features characteristic of SG can be seen viz. a lack of saturation (Fig. 2a) and an irreversible behavior in the isothermal magnetization curves, $M(H)$, (Fig. 2b) [17,18]. Concerning the former, (a) a difference between the curves measured in increasing and decreasing fields occurs, and (b) a lack of saturation of $M(H)$-curves exists. Using the extrapolated to $\mu_oH=0$ values of $M(H)$, a magnetic moment per atom was evaluated to range between ~0.25$\mu_B$ at $x$=45 to ~0.1$\mu_B$ at $x$=53.

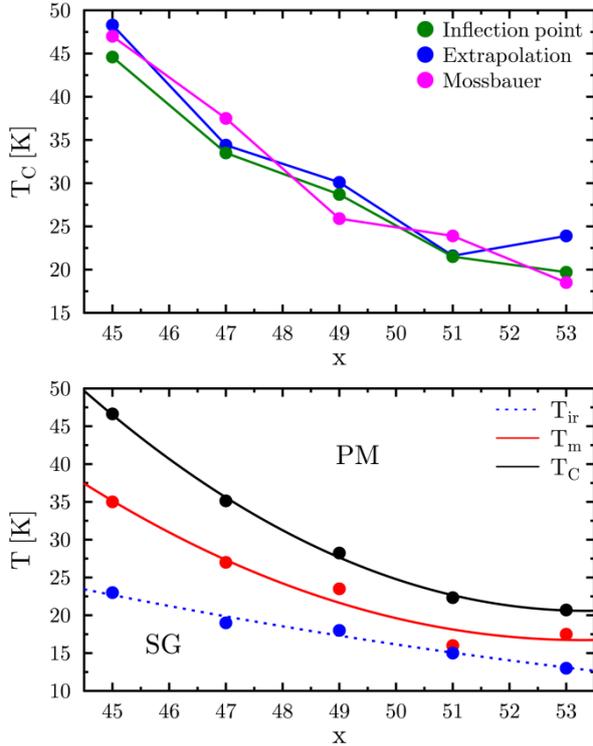

Further evidence in favor of SG as a magnetic ground state in our samples is the cusp in the ac susceptibility at a well-defined temperature, $T_f$, that shifts with frequency, $f$, towards higher temperature – see Fig. 3. The figure of merit, the relative temperature shift per frequency decade, $RTS = \Delta T_f / [T_f \Delta \log_{10}(f)]$, is equal to 0.012. Such a small value is characteristic of metallic spin glasses with RKKY interactions [19, 20].

Fig. 5. (Color online) The magnetic ordering temperature, $T_C$, as determined in 3 ways described in the legend vs. Mo content, $x$, (upper panel) and tentative magnetic phase diagram of the σ-Fe$_{100-x}$Mo$_x$ alloys (lower panel)

The magnetic ordering temperature, $T_C$, was determined in three ways viz. from the: (1) inflection point of the $M(T)$-curves, (2) extrapolation of a linear part of the $M(T)$-curves, and (3) temperature dependence of the average hyperfine field. Thus obtained $T_C$-values are displayed in Fig. 5. They are consistent with each other and range between ~46 K for $x$=45 and ~22K for $x$=53. Noteworthy, they are close to those found for σ-FeCr compounds [7].

Based on the characteristic temperatures determined in this study, a tentative magnetic phase diagram of the σ-FeMo alloy system could have been proposed – see Fig. 5.

Whatever the type of the magnetic ordering prior to the *SG* state is (it can be hopefully elucidated using neutron-diffraction measurements), the data shown in Figs 1, 2 and 3 give clear evidence that the ground magnetic state of the investigated samples is constituted by SG.

**Acknowledgements**

This work was supported by The Ministry of Science and Higher Education of the Polish Government and by The National Science Center (Grant DEC-2012/05/B/ST3/03241).